\documentclass{article}
\usepackage{graphicx}
\usepackage{amsmath,amssymb,amsfonts}
\newtheorem{theorem}{Theorem}[section]

\newtheorem{corollary}[theorem]{Corollary}
\title{Black Hole Interaction Energy}
\author{Sergio Dain\\
  Max-Planck-Institut f\"ur Gravitationsphysik\\
  Am M\"uhlenberg 1\\
  14476 Golm\\
  Germany}

\begin{document}
\maketitle

\begin{abstract}
  The interaction energy between two black holes at large separation
  distance is calculated.  The first term in the expansion corresponds
  to the Newtonian interaction between the masses. The second term
  corresponds to the spin-spin interaction. The calculation is based
  on the interaction energy defined on the two black holes initial
  data. No test particle approximation is used. The relation between
  this formula and cosmic censorship is discussed.
\end{abstract}

\section{Introduction}
The purpose of this article is to prove, under appropriate
assumptions, the following statement: the interaction energy, at large
separation distance $l$, between two black holes of masses $M_1$, $M_2$
and spins $J_1$, $J_2$ is given by
\begin{equation}
  \label{eq:32}
  E=\frac{-M_1M_2}{l}+\frac{-J_1\cdot J_2+3(J_1\cdot\hat n)(J_2\cdot\hat
  n)}{l^3} + \text{ higher order terms},
\end{equation}
where $\hat n$ is a unit  vector which points inward along the line
connecting the black holes. Before giving a precise
definition of the parameters involved in Eq. (\ref{eq:32}), I want to
discuss its physical meaning.

The first term in Eq. (\ref{eq:32}) has the Newtonian form. For two
point particles of masses $M_1$ and $M_2$ separated by an Euclidean
distance $l$, the Newtonian interaction energy between them is given by
$-M_1M_2/l$. The fact that this term appears also for a two black
holes system in General Relativity can be expected from the weak field
limit of Einstein's equations. The second term in Eq. (\ref{eq:32}),
which involves the spins, has analogous form to the dipole-dipole
electromagnetic interaction; there exists an analogy between magnetic
dipole in electromagnetism and spin in general relativity (see
\cite{Wald72}). In the electromagnetic case, $l$ is the Euclidean
distance between two charge distributions and $J_1$, $J_2$ the
corresponding dipole moments of them.  However, the gravitational
black hole spin-spin interaction has the opposite sign to the
electromagnetic one.  The first evidence of this fact was given by
Hawking \cite{Hawking72}.  I want to reproduce Hawking's argument here
because it points out the connection between Eq. (\ref{eq:32}) and the
cosmic censorship conjecture (see also the discussion in
\cite{Wald72}). In the argument, we  assume the following two
consequences of  weak cosmic censorship and the theory of black
holes (cf. \cite{Hawking73} \cite{Wald84}, see also \cite{Wald99}):

\begin{itemize}
\item[(i)] Every apparent
  horizon must be entirely contained within the black hole event horizon.

\item[(ii)] If matter satisfies
  the null energy condition (i.e. if $T_{ab}k^ak^b\geq 0$ for all null 
  $k^a$), then  the area of the event horizon of a black hole cannot
  decrease in time.
\end{itemize}
We also assume:
\begin{itemize}
\item[(iii)] All black holes eventually settle down to a final Kerr
  black hole.
 
\end{itemize}

Consider a system of two black holes such that, at a given time, the
separation distance between them is large. Then, there must exist a
Cauchy surface in the asymptotically flat region of the space time
such that the intersection of the hypersurface with the event horizon
has two disconnected component of areas $A_1$ and $A_2$. Since the
black holes are far apart, these areas can be approximated by the Kerr
formula
\begin{equation}
  \label{eq:26}
  A_1=8\pi \left (M^2_1+\sqrt{M_1^4-J_1^2}\right ), \quad  A_2=8\pi
  \left(M^2_2+\sqrt{M_2^4-J_2^2}\right). 
\end{equation}
At late times, after the collision, the system will settle down to a
Kerr black hole. Hence, there must exist another Cauchy hypersurface
such that its  intersection with  the event horizon will have area
\begin{equation}
  \label{eq:27}
  A_f=8\pi \left(M^2_f+\sqrt{M_f^4-J_f^2}\right),
\end{equation}
where $M_f$ is mass of the final black hole and $J_f$ is its
final angular momentum. By (ii) we have
\begin{equation}
  \label{eq:28}
  A_f\geq A_1+A_2.
\end{equation}
Since gravitational waves have positive mass, we also have
\begin{equation}
  \label{eq:29}
  M_f\leq M_1+M_2.
\end{equation}
In general, gravitational waves will carry angular momentum. But in
\emph{axially symmetric} space-times the total angular momentum is a
conserved quantity, since it can be defined by a Komar integral (cf.
\cite{Komar59} and also \cite{Wald84}). Then, in this case we have
\begin{equation}
  \label{eq:30}
  J_f=J_1+J_2.
\end{equation}
Using Eqs. (\ref{eq:26}), (\ref{eq:27}), (\ref{eq:28}) and
(\ref{eq:30}) it is possible to obtain an upper bound, which depends
on $J_1$ and $J_2$, to the total amount of radiation emitted by the
system $M_1+M_2-M_f$. It can be seen that if $J_1$ and $J_2$ have the
same sign, this upper bound is smaller than if they have opposite sign.
This suggests that there may be a spin-spin force between the black
holes that is attractive if the angular momentum have opposite
directions and repulsive if they have the same direction. Presumably,
in the second case the system expends energy in doing work against the
spin repulsive force, and for this reason this energy is not available
to be radiated via gravitational radiation.

Hawking's argument only suggests that the spin interaction energy
between black holes has in fact this sign dependence with respect to
the spins. It is not a proof, first because there is no proof for the
weak cosmic censorship conjecture (i)-(ii) and for the assumption
(iii).  Second, because even if we
assume (i)--(iii) the argument only shows that an upper bound of the
total amount of radiated energy has this sign dependence in terms of
$J_1$ and $J_2$, but the real amount of gravitational radiation can,
in principle, have other dependence.  In fact, the total amount of
gravitational radiation produced by such systems, as numerical studies
show, is much smaller than this bound. This upper bound is $50\%$ of
the total mass when the spins are antiparallel,  the black holes are
extreme ($J^2=M$), and have equal masses; when the spins are zero or
when the black holes are extreme with parallel spins, the upper bound
is $29\%$ of the total mass. On the other hand, in the numerical
calculations the maximum amount of radiation emitted by this type of
system is about $3\%$ of the total mass, see
\cite{Baker01b} \cite{Baker02} for a recent calculation and also
\cite{Lehner01} for an up to date review on the subject.
However, the numerical studies show that the system indeed
radiates less when the spins are parallel than when they are
antiparallel. Moreover,  Wald \cite{Wald72} proves
that the interaction energy between a test particle with spin $J_2$
and a stationary background of spin $J_1$ has precisely this sign
dependence.  Wald shows that the spin-spin interaction energy
has the form
\begin{equation}
  \label{eq:31}
\frac{-J_1\cdot J_2+3(J_1\cdot\hat n)(J_1\cdot\hat n)}{l^3},  
\end{equation}
where $l$ and $\hat n$ are defined as follows. The stationary field is
expanded at large distance with respect to Cartesian asymptotic
coordinates $x^i$,  here $l$ is the Euclidean radius with respect to
$x^i$ and $\hat n^i=x^i/l$. Eq. (\ref{eq:31}) has been also proved by
D'Eath using post-Newtonian expansions \cite{death75}. It is important to
note that Eq. (\ref{eq:31}) gives an indirect evidence in support of
(i)-(iii).

In this article I want to prove Eq. (\ref{eq:31}) without using either 
the  particle or post-newtonian approximation. The proof is based
on an interaction energy defined on the two black hole initial
data. This interaction energy is genuinely non linear; it does not
involve any approximation. 

The plan of the paper is as follows. In section \ref{main} the main
results are given. In section \ref{by} theorem
\ref{T1} is proved; in section \ref{twokerr} we prove corollary
\ref{C1}. Finally, in section \ref{dis} an alternative definition of
the interaction energy is discussed.

\section{Main Result} \label{main}

The strategy I will follow was given by Brill and Lindquist
\cite{Brill63}. It is based on the analysis of \emph{initial data set}
with many \emph{asymptotic ends}. An \emph{initial data set} for the
Einstein vacuum equations is given by a triple $(\tilde S, \tilde
h_{ab}, \tilde K_{ab})$ where $\tilde S$ is a connected
3-dimensional manifold, $\tilde h_{ab} $ a (positive definite)
Riemannian metric, and $\tilde K_{ab}$ a symmetric tensor field on
$\tilde S$. They satisfy the vacuum constraint equations
\begin{equation}
 \label{const1}
\tilde D^b \tilde K_{ab} -\tilde D_a \tilde K=0,
\end{equation}
\begin{equation}
 \label{const2}
\tilde R + \tilde K^2-\tilde K_{ab} \tilde K^{ab}=0,
\end{equation}
on $\tilde S$, where $\tilde D_a$ is the covariant derivative with
respect to $\tilde h_{ab}$, $\tilde R$ is the trace of the
corresponding Ricci tensor, $\tilde K=\tilde h^{ab} \tilde K_{ab}$,
and $a,b,c...$ denote abstract indices.  Tensor indices of quantities
with tilde will be moved with the metric $\tilde h_{ab}$ and its
inverse $\tilde h^{ab}$. The data will be called \emph{asymptotically
  flat} with $N+1$ \emph{asymptotic ends}, if for some compact set
$\Omega$ we have that $\tilde S\setminus \Omega =\sum_{k=0}^N\tilde
S_k$, where $\tilde S_k$ are open sets such that each $\tilde S_k$ can
be mapped by a coordinate system $x^j$ diffeomorphically onto the
complement of a closed ball in $\mathbb{R}^3$ such that we have in
these coordinates
\begin{equation} 
\label{pf1}
\tilde h_{ij}=(1+\frac{2M_k}{r})\delta_{ij}+O(r^{-2}),
\end{equation}
\begin{equation} 
\label{pf2}
\tilde K_{ij}=O(r^{-2}),
\end{equation}
as $r= ( \sum_{j=1}^3 ({x^j})^2 ) ^{1/2} \to \infty$ in each set
$\tilde S_k$; where $i,j \cdots$, which take values $1, 2, 3$, denote
coordinates indices with respect to the given coordinate system $x^j$,
and $\delta_{ij}$ denotes the flat metric.  We will call the
coordinate system $x^i$ an \emph{asymptotic coordinate system} at the
end $k$.  Each asymptotic region $\tilde S_k$ has a different
asymptotic coordinate system.  The constant $M_k$ denotes the ADM
mass\cite{Arnowitt62} of the data at the end $k$.  These conditions
guarantee that the mass, the linear momentum, and the angular momentum
of the initial data set are well defined at every end.

For $N\geq 1$, this class of data contains, in general, apparent
horizons. The existence of apparent horizons leads us to interpret
these data as representing initial data for black-holes. Their evolution
 will presumably contain an event horizon, according to the
standard theory of black holes \cite{Hawking73}.  The validity of this
picture depends, of course, on the cosmic censorship conjecture. The
only statement about the evolution of the data that we can make is the
geodesic incompleteness of the space time. In general, in order to
prove the geodesic incompleteness of a space time, one needs to know
that the data contain a trapped surface in order to apply the
singularities theorems \cite{Hawking73}. However, in this particular
case, since the topology of the data is not trivial, the geodesic
incompleteness of the space time follows directly from a theorem
proved by Gannon \cite{gannon75}.

\begin{figure}[htbp]
  \begin{center}
    \includegraphics[width=6cm]{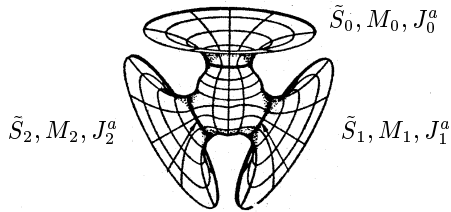}
    \caption{Initial data with three asymptotic ends ($N=2$). For  each
      asymptotic region $\tilde S_k$, we have the corresponding mass
      $M_k$ and total angular momentum $J^a_k$.}
    \label{fig:1}
  \end{center}
\end{figure}

For simplicity  we will fix $N=2$, see Fig. \ref{fig:1}. 
In this case the data can be interpreted as initial data with two
black holes. This interpretation is suggested by the following fact:
when an appropriate distance parameter is large compared with the
masses $M_k$, then it can be seen numerically that only two
disconnected apparent horizons appear. For time symmetric data, these
numeric calculations have been done in \cite{Brill63}; the non-time
symmetric case has been studied by Cook (see \cite{Cook00} and
references therein).  It is not clear that the number of apparent
horizons is the number of black holes contained in the data, since
even when there are two disconnected apparent horizons, the intersection
of the event horizon with the initial data can be connected. However,
at large separation distance, this seems to be a reasonable assumption,
which is confirmed by the numerical evolutions \cite{Lehner01}.

\begin{figure}[htbp]
  \begin{center}
    \includegraphics[width=6cm]{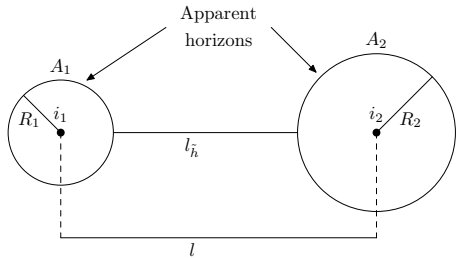}
    \caption{An initial data set with three asymptotic end points and  only 
      two disconnected apparent horizons of area $A_1$, $A_2$, and
      radii $R_1$, $R_2$. The points $i_1$ and $i_2$ represent the
      two other infinities $1$ and $2$. The geodesic distance
      $l_{\tilde h}$ is computed with the physical
      metric $\tilde h_{ab}$. The parameter $l$ is computed with the conformal
      metric. The geodesic distance between $i_1$ and $i_2$
      with respect to the physical metric is infinite.}
    \label{fig:2}
  \end{center}
\end{figure}

Brill and Lindquist define the following interaction energy at the end 
$k$
\begin{equation}
  \label{eq:1}
  E_{k}=M_{k}-\sum^N_{\substack{k'=0\\ k'\neq k}}M_{{k'}}.
\end{equation}
The energy $E_k$ is a geometric quantity; its definition does not
involves any approximation. The question now is how to calculate
$E_{k}$ in terms of physically relevant parameters. The first problem is
how to define an appropriate separation distance between the black
holes.  When there are two apparent horizons, there is a well defined
separation distance $l_{\tilde h}$ defined as the minimum geodesic
distance between any two points in the two different horizons, see
Fig. \ref{fig:2}.  

However, the distance $l_{\tilde h}$ is hard to
compute. The location of the apparent horizons can be calculated only
numerically.  Since we are only interested in the energy at large
separations, instead of $l_{\tilde h}$ we will use another parameter
$l$, and we will argue that $l_{\tilde h}\approx l$ in this limit. The
definition of the parameter $l$ is related to the way in which one can
construct solutions of the constraint equations with many asymptotic
ends. The conformal method (cf.  \cite{Choquet99}, \cite{Choquet80}
and the references therein) is a general method for constructing
solutions of the constraint equations.  We assume that
$h_{ab}$ is a positive definite metric with covariant derivative
$D_a$, and $K^{ab}$ is a trace-free (with respect to $h_{ab}$),
symmetric tensor, satisfying
\begin{equation} 
\label{diver}
D_a K^{ab}=0 \quad\mbox{on}\quad \tilde S.
\end{equation}
Let $\varphi$ be a solution of 
\begin{equation} 
\label{Lich}
L_h \varphi=-\frac{1}{8}K_{ab}K^{ab}\varphi^{-7}
\quad \mbox{on}\quad \tilde S , 
\end{equation}
where $L_h=D^aD_a-R/8$ and $R$ is the scalar curvature of the metric
$h_{ab}$.  Then the physical fields $(\tilde h, \tilde K)$ defined by
$\tilde{h}_{ab} = \varphi^4 h_{ab}$ and $\tilde{K}^{ab} =
\varphi^{-10}K^{ab}$ will satisfy the vacuum constraint equations on
$\tilde S$. We have assumed that $K^{ab}$ is trace-free; hence
$\tilde K^{ab}$ will be also trace-free with respect to $\tilde
h_{ab}$. That is, the initial data set will be \emph{maximal}.
 
To ensure asymptotic flatness of the data at the each end $k$, we will
require the following boundary conditions. Let $i_1$ and $i_2$ be two,
arbitrary points in $\mathbb{R}^3$, with coordinates $x_1^j$ and
$x_2^j$ in some Cartesian coordinate system $x^i$. Define the manifold 
$\tilde S$ by  $\tilde S= \mathbb{R}^3\setminus\{i_1,i_2\}$. Assume
that $h_{ab}$ is regular on $\mathbb{R}^3$. At infinity we will impose 
  the following fall
off behavior
\begin{equation}
  \label{eq:33}
  h_{ij}=\delta_{ij} +  O(r^{-2}),
\end{equation}
\begin{equation}
  \label{eq:42}
K^{ab} = O(r^{-2}),
\end{equation}
\begin{equation}
  \label{eq:37}
   \varphi =1+ O(r^{-1}).
\end{equation}
At the points $i_1$ and $i_2$ we require  
\begin{equation} 
\label{Psii}
K^{ab} = O(r_1^{-4}), \quad K^{ab} = O(r_2^{-4}),  
\end{equation}
where
\begin{equation}
  \label{eq:7}
  r_1=\left(\sum_{i=1}^3(x^i-x_1^i)^2\right)^{1/2},\quad
r_2=\left(\sum_{i=1}^3(x^i-x_2^i)^2\right)^{1/2},
\end{equation}
and
\begin{equation}
  \label{eq:36}
 \lim_{r_1\rightarrow 0} r_1\varphi = \frac{m_1}{2}, \quad \lim_{r_2\rightarrow
   0} r_2\varphi = \frac{m_2}{2}, 
\end{equation}
where $m_1$ and $m_2$ are positive constants. Note that both $\varphi$
and $K^{ab}$ are singular at $i_1$, $i_2$. 

One can prove that the data so constructed will be asymptotically flat
at the three ends.  We have made an
artificial distinction between the end $0$, given by $r\rightarrow
\infty$, and the ends $1$ and $2$.  It is possible to discuss the same
construction in a more geometrical way, such that all ends are
treated equally; see \cite{Beig94}, \cite{Friedrich88},
\cite{Friedrich98}, \cite{Dain99b}.  However, since our final goal is
to calculate the interaction energy at one end, it is convenient to make
this distinction. The coordinate system $x^i$ and the corresponding
flat metric in the expansion Eq.  (\ref{eq:33}), gives the Euclidean
distance $l$ between $i_1$ and $i_2$
\begin{equation}
  \label{eq:9}
 l=\left(\sum_{i=1}^3(x_2^i-x_1^i)^2\right)^{1/2}, 
\end{equation}
which will be our separation distance parameter, see Fig. \ref{fig:2}.

In general, Eq.~(\ref{Lich}) is non-linear. However if we assume that
the data is time symmetric, i.e. $K^{ab}=0$, then it becomes a linear
equation for $\varphi$. If we assume that the conformal metric is
flat, we obtain a Laplace equation for $\varphi$. The solution  of
this equation that satisfies the boundary conditions
Eqs.~(\ref{eq:36}) and (\ref{eq:37}) is given by
\begin{equation}
  \label{eq:44}
  \varphi_0=1+\frac{m_1}{2r_1}+ \frac{m_2}{2r_2}.
\end{equation}
This solution was found by Brill and Lindquist in \cite{Brill63}. In
this case it is possible to calculate explicitly the interaction
energy (\ref{eq:1}) in terms of the masses and the separation
distance. The result is the following.
\begin{theorem}[Brill-Lindquist]
 \label{BL}
  Let $h_{ab}$ the flat metric and $\tilde K^{ab}=0$. Then the
  interaction energy defined by Eq. (\ref{eq:1}) is always
  negative. Moreover, when $l$ is large compared with $M_k$ the
  following expansion holds
  \begin{equation}
    E_0=-\frac{M_1M_2}{l}+\text{ higher order terms}.
  \end{equation}
\end{theorem}
Giulini \cite{Giulini90} has computed the higher order terms for these
data and other conformally flat time symmetric data with different
topologies. In those examples  the Newtonian term is invariant but
the higher order terms depend on the particular initial data.

In order to discuss  spin-spin interaction, we need initial data with
non trivial angular momentum, that is we have to allow for non trivial
extrinsic curvature in the data. At each end we have the angular
momentum $J_{k}$ given by
\begin{equation}
  \label{eq:34}
  J_{k}^a=\frac{1}{8\pi} \lim_{r\to \infty } 
\int_{S_{r}} \,r
K_{bc}\,n^b \epsilon^{cad}\,n_d\,dS_{r},
\end{equation}
where $S_r$ is a two sphere defined in the asymptotic region $\tilde
S_k$ and $n^a$ is its outward unit normal vector. In Eq. (\ref{eq:34}) we
can use either $K^{ab}$ or $\tilde K^{ab}$ because the conformal
factor satisfies (\ref{eq:37}). In general the angular momentum at
each end is \emph{not} determined by the intrinsic angular momentum of
each black hole. It includes also the angular momentum of the
gravitational field surrounding the black holes. Then, in general,
there is no relations between $J_0$, $J_1$ and $J_2$, these three
quantities can be freely prescribed. But in the presence of symmetries
these quantities can not be given freely any more. Moreover, in the
presence of \emph{conformal symmetries} of the metric there exists a well
defined quasilocal definition of angular momentum. Assume that
$\xi^a$ is a conformal Killing vector; that is, a solution of the
equation $(\mathcal{L}_h \xi)^{ab} =0$, where
\begin{equation}
  \label{eq:35}
  (\mathcal{L}_h \xi)^{ab} = D^a\xi^b+D^b\xi^a -
  \frac{2}{3}\,h^{ab}\,D_c \xi^c.
\end{equation}
If the initial data is \emph{maximal}, i.e., $K=0$, then the vector
$K^{ab}\xi_b$ is divergence free. Hence, for each conformal symmetry
$\xi^a$ we have the associated integral
\begin{equation}
  \label{eq:41}
  I_\xi=\int_S K^{ab}\xi_b n_a dS,
\end{equation}
where $S$ is a close 2-surface and $n^a$ its outward unit normal
vector. This integral is a conformal invariant. It can be calculated
also in terms of tilde quantities.  The integral (\ref{eq:41}) will be
non zero only if the vector $K^{ab}\xi_b$ is singular at some points;
in our case it will be singular at two points: the location of the
holes. Then the integral in Eq. \eqref{eq:41} will have three
different values depending on whether the surface $S$ encloses one
hole, two holes or no hole. In the later case, $I_\xi=0$. If we chose
$\xi^a$ to be a rotation, $I_\xi$ will gives the corresponding
component of the quasilocal angular momentum. If the data is
conformally flat we have $10$ conformal Killing vectors. In
particular, we have the three rotations and hence the complete
definition of quasilocal angular momentum.  These quantities will be
defined only on this slice and will generally not be preserved in the
evolution. They will be only preserved if the space time admits a
Killing vector. In this case they will coincide with the corresponding
Komar integral. The space time will admit a Killing vector field if
$\xi^a$ is a Killing vector for the whole initial data; that is,
$\pounds_\xi \tilde h_{ab}= \pounds_\xi \tilde K^{ab}=0$, where
$\pounds_\xi$ is the Lie derivative with respect to $\xi^a$.  A
conformally flat, maximal, slice can be interpreted as a instant of
time in which the gravitational field carries no angular momentum and
no linear momentum itself, and hence these quantities are carried only
by the ``sources'', which in this case are the black holes. Data
containing matter with compact support can be also constructed.

There exist in the literature other definitions of quasilocal angular
momentum (\cite{Penrose83} \cite{Ludvigsen83}
\cite{Dougan91}\cite{Szabados96}), which are applicable for an
arbitrary closed 2-surfaces in the spacetime. It is not  clear if
any of these definitions will agree with Eq. (\ref{eq:41}) in the
particular case of 2-surface lying on a conformally flat
3-hypersurface. 

From the discussion above, we conclude that in the case of conformally
flat, maximal data we have,
\begin{equation}
  \label{eq:43}
  J^a_0+J^a_1+J^a_2=0.
\end{equation}
For an observer placed in the asymptotic end $0$ the system will look
like two black hole with spins $-J^a_1$ and $-J^a_2$, and the total
angular momentum will be $J^a_0=-J^a_1-J^a_2$. For a more general
discussion of conformal symmetries on initial data see \cite{Beig96}
and \cite{Dain99}; in particular in those articles a generalization of
Eq. (\ref{eq:43}) which includes linear momentum is proved.

Bowen and York obtain a simple model for a conformally flat data set 
which represents two black holes with spins \cite{Bowen80}.  Brandt and
Br\"ugmann \cite{Brandt97b} study these data with the $N+1$
asymptotic ends boundary conditions given by Eqs. (\ref{Psii}),
(\ref{eq:42}), (\ref{eq:36}) and (\ref{eq:37}).  For these data, the
conformal second fundamental form is given by
\begin{equation}
 \label{eq:kby}
K^{ab}=K_1^{ab}+K_2^{ab}, 
\end{equation}
where 
\begin{equation}
  \label{eq:6}
  K_1^{ab}=\frac{6}{r_1^3}n_1^{(a}\epsilon^{b)cd}J_{1\,c}n_{1\,d},\quad  
K_2^{ab}=\frac{6}{r_2^3}n_2^{(a}\epsilon^{b)cd}J_{2\,c}n_{2\,d},
\end{equation}
and 
\begin{equation}
  \label{eq:8}
  n_1^i=\frac{x^i-x_1^i}{r_1},\quad 
  n_2^i=\frac{x^i-x_2^i}{r_2},
\end{equation}
where $J_1^c$, $J_2^c$ are constants and $\epsilon^{bcd}$ is the flat
volume element. One can check that the constants $J_1^c$ and $J_2^c$
give the the angular momentum at the ends $1$ and $2$ respectively.
The tensors \eqref{eq:6} are divergence free and trace free with respect to
the flat metric in $\mathbb{R}^3\setminus\{i_1,i_2\}$.
 
The first result of this article is the following theorem. 
\begin{theorem}\label{T1}
  Let $h_{ab}$ be the flat metric and $K^{ab}$ be given by Eqn.
  (\ref{eq:kby}). Then, the interaction energy (\ref{eq:1}) is given
  by
  \begin{equation}
    \label{eq:38}
    E_0=\frac{-M_1M_2}{l}+\frac{-J_1\cdot J_2+3(J_1\cdot\hat n)(J_1\cdot\hat
  n)}{l^3} + \text{ higher order terms}.
  \end{equation}
\end{theorem}
The expansion of the dimensionless quantity $E_0/l$ is made in terms
of the dimensionless parameters $M_1/l$, $M_2/l$, $J_1/l^2$  and
$J_2/l^2$. By higher order terms we mean terms of cubic order in those
parameters. We prove Theorem \ref{T1}  in section \ref{by}. 

Using similar arguments, Gibbons\cite{Gibbons02}\cite{Hawking73b} has
obtained the qualitative sign dependence of the spin-spin interaction
for a certain class of axially symmetric, conformally flat, data. Note
that in theorem \ref{T1}, the data are non axially symmetric in
general, since the spins can point in arbitrary directions. Bonnor
\cite{Bonnor02} has studied the spin-spin interaction using an exact,
axially symmetric solution of the Einstein-Maxwell equations, his
result also qualitatively agrees with the spin-spin term in Eq.
(\ref{eq:38}).

The Bowen-York data are, of course, very special. The natural question
is how to generalize theorem \ref{T1} for more general data.  For
general asymptotically flat data with three asymptotic ends we can not
even expect to recover the Newtonian interaction term.  Take for
example a time symmetric initial data with only two ends.  Choosing
the conformal metric appropriately, one can easily construct data such
that the difference $M_1-M_0$ is arbitrary. That means that we are
putting more radiation in one end than in the other. If we add a third
end with small mass, the new interaction energy will be dominated by
the difference $M_1-M_0$ and hence will be not related with any
Newtonian force.  Hence, theorem \ref{BL} is not true for general
asymptotically flat metrics with many asymptotic ends.

The interaction energy defined by (\ref{eq:1}) can have the meaning of
a two black holes interaction energy only if is it possible to
distinguish in the data two objects that are similar to the Kerr black
hole when the separation distance is large. I will call this class of
data  \emph{two Kerr-like black hole initial data}. The existence of
these data has been proved in \cite{Dain00c} \cite{Dain99b}. The
following two properties of the Kerr initial data, in the standard
Boyer-Lindquist coordinates, are important: i) The data are
conformally flat up to order $O(J^2)$ ii) The leading order term of
the second fundamental form is given by the Bowen-York one
(\ref{eq:6}). Then one can expect that Eq.  (\ref{eq:38}) is unchanged
in the principal terms for this class of data. 

More precisely, a \emph{two Kerr-like black hole initial data} can be
constructed as follows (see \cite{Dain00c} and \cite{Dain99b} for
details). Take a slice of the Kerr metric, with parameters $M_{K_1}$
and $J_1$, in the standard Boyer-Lindquist coordinates. Choosing the
appropriate conformal factor, the conformal metric can be written in
the following form
\begin{equation}
  \label{eq:39b}
  h^{K_1}_{ab}=\delta_{ab}+ h^{R_1}_{ab},
\end{equation}
where $ h^{R_1}_{ab}=O(J_1^2)$. In the same way the conformal second
fundamental form can be written as
\begin{equation}
  \label{eq:40b}
  K_{K_1}^{ab}=K_1^{ab}+K^{ab}_{R_1},
\end{equation}
where $ K^{ab}_{R_1}=O(J_1^2)$ and $K_1^{ab}$ is given by
(\ref{eq:6}).  Take another Kerr metric, with parameters $M_{K_2}$ and
$J_2$, and define the following conformal metric
\begin{equation}
  \label{eq:39}
  h^{KK}_{ab}=\delta_{ab}+ h^{R_1}_{ab}+h^{R_2}_{ab},
\end{equation}
and the following conformal second fundamental form
\begin{equation}
  \label{eq:40}
  K^{ab}_{KK}=\bar K_{K_1}^{ab}+\bar
  K_{K_2}^{ab}+(\mathcal{L}_{h_{KK}} w)^{ab},  
\end{equation}
where the bar means the trace free part of the tensor with respect to
the metric (\ref{eq:39}) and $w^a$ is chosen such that $K^{ab}_{KK}$
is divergence free and trace free with respect to the metric (\ref{eq:39}).
In \cite{Dain00c} \cite{Dain99b} it has been proved that such vector
$w^a$ exists and is unique. Using (\ref{eq:39}) and (\ref{eq:40}),
solve Eq. (\ref{Lich}) with the boundary conditions (\ref{eq:36}) and
(\ref{eq:37}) where
\begin{equation}
 \label{eq:mmk}
m_1=\sqrt{M^2_{K_1}-J^2_1/M^2_{K_1}}, \quad
m_2=\sqrt{M^2_{K_2}-J^2_2/M^2_{K_2}}. 
\end{equation}
The existence of a unique solution has been proved in \cite{Dain00c}
\cite{Dain99b}. 

If we chose $J^a_1$ and $J^a_2$ to point in the same direction, the
data will be axially symmetric. In this particular case, we can use
the integral (\ref{eq:41}) to calculate the quasilocal angular
momentum of each of the black holes. The result will be $J^a_1$ and
$J^a_2$. However, in general, this class of data will admit no
conformal Killing vector.  In this general situation, it is very hard
to compute the quasilocal spins of each of the black holes. However,
when the separation distance is large, $J^a_1$ and $J^a_2$ will give
approximately the angular momentum of each of the black holes because
this class of data have a far limit to the Kerr initial data.  In
other words, $J^a_1$ and $J^a_2$ give the spins of one black hole when
the parameters of the other are set to zero.

For this class of data we have the following result. 
\begin{corollary}
 \label{C1}
 For the two Kerr-like data defined above, the formula (\ref{eq:38})
 for the interaction energy holds.
\end{corollary}
We prove this Corollary in section \ref{twokerr}. 

\section{Interaction Energy for the spinning Bowen-York initial data}\label{by}
In this section we will prove theorem \ref{T1}. For Bowen-York data,
with the boundary condition (\ref{eq:36}) and (\ref{eq:37}), Eq.
(\ref{Lich}) for the conformal factor can be written in the following
form\cite{Brandt97b}
\begin{equation}
  \label{eq:12}
  \Delta u = -\frac{K^{ab}K_{ab}}{8\varphi^7}, \quad
  \varphi=\varphi_0+u, 
\end{equation}
with the boundary condition
\begin{equation}
  \label{eq:13}
  \lim_{r\rightarrow \infty} u =0,
\end{equation}
where $\varphi_0$ is defined in Eq. (\ref{eq:44}), with $m_1$ and
$m_2$ arbitrary positives constants, and $K^{ab}=K_1^{ab}+K_2^{ab}$ is
given by (\ref{eq:6}).

The coordinates $x^i$ are asymptotic coordinates for the end $0$. The
total mass at $0$ is given by
\begin{equation}
  \label{eq:15}
  M_0=m_1+m_2+2u_\infty,
\end{equation}
where $u_\infty$ is the term which goes like $1/r$ in the solution $u$
of Eq. (\ref{eq:12}) and is given by 
\begin{equation}
  \label{eq:16}
  u_\infty=\frac{1}{4\pi}\int_{\mathbb{R}^3} \frac{K^{ab}K_{ab}}{8\varphi^7}dx^3.
\end{equation}
We want to calculate the masses $M_1$ and $M_2$ for the other ends.
The asymptotic coordinates for the other ends are
\begin{equation}
  \label{eq:17}
  \hat x_{i_1}^i=\frac{m_1^2}{4}\frac{(x^i-x_1^i)}{r_1^2}, \quad  \hat
  x_{i_2}^i=\frac{m_2^2}{4}\frac{(x^i-x_2^i)}{r_2^2}. 
\end{equation}
Take for example the end point $i_1$, in $\hat x_{i_1}^i$ coordinates
we have
\begin{equation}
  \label{eq:18}
  \tilde h_{\hat i \hat j}=\left(1+\frac{m_1}{2\hat
    r_1}+\frac{m_1m_2}{4l\hat
    r_1}+\frac{u(i_1)}{2\hat{r_1}}\right)\delta_{\hat i \hat j} +
O(1/\hat r^2_1), 
\end{equation}
where we have used
\begin{equation}
  \label{eq:19}
  r_2=l+O(1/\hat r_1),
\end{equation}
and $\hat r_1$ is the Euclidean radius with respect to $ \hat
x_{i_1}^i$. 
Then the mass at this end is given by  
\begin{equation}
  \label{eq:20}
  M_1 =m_1\left(1+\frac{m_2}{2l}+u(i_1) \right),
 \end{equation}
where $u(i_1)$ denote the value of the function $u$ at the point $i_1$.
In analogous way we obtain the mass at the end $2$
\begin{equation}
  \label{eq:20b}
 M_2 =m_2\left(1+\frac{m_1}{2l}+u(i_2) \right).
\end{equation}

Then the interaction energy  at the end $i_0$ is given by
\begin{equation}
  \label{eq:22}
  E_0=M_0-M_1-M_2=-\frac{m_1m_2}{l}+2u_\infty-m_1u(i_1)-m_2u(i_2).
\end{equation}
Using Eq. (\ref{eq:12}) and the Green's function   for the Laplacian, we
obtain the following integral representation of the terms involving
$u$ in Eq. (\ref{eq:22})
\begin{equation}
  \label{eq:10}
2u_\infty-m_1u(i_1)-m_2u(i_2)=\frac{1}{16\pi}\int_{\mathbb{R}^3}
\frac{K^{ab}K_{ab}}{\varphi^7}\left(1-\frac{m_1}{2r_1}-
  \frac{m_2}{2r_2}\right)dx^3.    
\end{equation}
The formula (\ref{eq:10}) involves the unknown function $u$. Using the
fact that $u$ is $O(J^2)$, we make an expansion of this integral in
terms of the parameters $J_1/l^2$, $J_2/l^2$, $m_1/l$ and $m_2/l$. We
obtain that the first non trivial term is given by
\begin{equation}
  \label{eq:11}
  2u_\infty-m_1u(i_1)-m_2u(i_2)\approx E^s,
\end{equation}
where
\begin{equation}
  \label{eq:23}
  E^s=\frac{1}{8\pi}\int_{\mathbb{R}^3} K_1^{ab}K_{2\,ab}\,  dx.
\end{equation}
The interaction energy, up to this order, is given by
\begin{equation}
  \label{eq:4}
  E=-\frac{M_1M_2}{l}+E^s,
\end{equation}
where we have used Eqs. (\ref{eq:20}) and (\ref{eq:20b}) to replace
$m_k$ by $M_k$, since up to this order they are equal.

All that remains is to compute the integral $E^s$.  This integral can,
in principle, be calculated explictly from the expressions
\eqref{eq:6}. However such a calculation is very complicated. Instead
of this we will calculate \eqref{eq:23} in the following way. The
tensors \eqref{eq:6} can be written like
\begin{equation}
  \label{eq:24}
{(\mathcal{L}_{\delta} v_1)}^{ab}=K_1^{ab},\quad (\mathcal{L}_{\delta}
v_2)^{ab}=K_2^{ab}   
\end{equation}
where $\mathcal{L}_{\delta}$ is the conformal Killing operator defined in
Eq. (\ref{eq:35}) for the flat metric, and
\begin{align}
  \label{eq:25}
  v_1^i & =-\epsilon^{ijk}J_{1\,j}n_{1\,k} r_1^{-2},\\
v_2^i & =-\epsilon^{ijk}J_{2\,j}n_{2\,k} r_2^{-2},
\end{align}
Let $B_{\epsilon_1}$ and $B_{\epsilon_2}$ be small balls centered at
$i_1$ and $i_2$ respectively, of radii $\epsilon_1$ and
$\epsilon_2$. We have that
\begin{equation}
  \label{eq:2}
  E^s=\frac{1}{8\pi}\lim_{\epsilon_1, \epsilon_2 \rightarrow 0}
  \int_{\mathbb{R}^3-B_{\epsilon_1}-B_{\epsilon_2}} K_1^{ab}K_{2\,ab}\,  dx.
\end{equation}
Using the Gauss theorem in
$\mathbb{R}^3-B_{\epsilon_1}-B_{\epsilon_2}$ we obtain
\begin{equation}
  \label{eq:3}
 \int_{\mathbb{R}^3-B_{\epsilon_1}-B_{\epsilon_2}} K_1^{ab}K_{2\,ab}\,
 dx= -2\int_{\partial B_{\epsilon_1}}  K_2^{ab}v_{1\,b}n_{1\,a}\,
 dS_{\epsilon_1} -2\int_{\partial B_{\epsilon_2}}  K_2^{ab}v_{1\,b}n_{2\,a}\,
 dS_{\epsilon_2},  
\end{equation}
where $n^a_1$ and $n^a_2$ are the outward normals to the two
surfaces $B_{\epsilon_1}$ and $B_{\epsilon_2}$.  In the limit
$\epsilon_1 \rightarrow 0 $ the first integral vanishes. 
We use here that $K_2^{ab}$ is regular in $B_{\epsilon_1}$.
The second integral, in the limit $\epsilon_2 \rightarrow 0 $ can be easily
calculated. We obtain 
\begin{equation}
  \label{eq:5}
  E^s=\frac{-J_1\cdot J_2+3(J_1\cdot\hat n)(J_1\cdot\hat
  n)}{l^3}.
\end{equation}

\section{Interaction energy for the two Kerr-like black holes initial
  data}\label{twokerr} 

The two Kerr-like black hole initial data set are solutions of the
following equation
\begin{equation} 
\label{Lich2kerr}
L_{h^{KK}} \varphi_{KK}=-\frac{K_{KK \,ab}K_{KK}^{ab}}{8\varphi_{KK}^{7}}
\quad \mbox{on}\quad \tilde S , 
\end{equation}
with the boundary conditions (\ref{eq:36}) and (\ref{eq:37}), where
$h^{KK}_{ab}$ and $K^{ab}_{KK}$ are given by (\ref{eq:39}),
(\ref{eq:40}) and $m_1$, $m_2$ are given by (\ref{eq:mmk}) in term of
the Kerr parameters.

The conformal factor for the Kerr initial data, with parameters $M_{K_1},
J_1$  can be written in the following form
\begin{equation}
 \label{eq:phik}
\varphi_{K_1}=1+\frac{m_1}{2r_1}+\varphi_{R_1},
\end{equation}
where $\varphi_{R_1}=O(J_1^2)$. We can decompose the two Kerr solution
in the following form
\begin{equation}
\varphi_{KK}=\varphi_0+\varphi_{R_1}+\varphi_{R_2}+u.
\end{equation}
Using that (\ref{eq:phik}) is a solution for one Kerr initial data, we
obtain that the first term in the expansion in $J_1$ and $J_2$ of the
function $u$ satisfies the following linear equation
\begin{equation}
\Delta u \approx -\frac{K_1^{ab}K_{2\,ab}}{8\varphi^7_0}. 
\end{equation}
Hence, the spin-spin interaction term has the same form as the
Bowen-York one. In an analogous way to the previous section, we obtain
that the interaction energy is given by
\begin{equation}
E_0 \approx  -\frac{M_1M_2}{l}+E^s,
\end{equation}
where $E^s$ is given by Eq. (\ref{eq:23}).

\section{Discussion}\label{dis}

We have shown, using the interaction energy defined by
Eq.~(\ref{eq:1}), that the spin spin interaction between black holes
of arbitrary masses and spins has an expansion of the form
(\ref{eq:38}).  This formula has been previously derived using a test
particle approximation \cite{Wald72} and post-newtonian expansions
\cite{death75}. The main improvement of the present calculation,
beside its simplicity, is that no approximation is used in the
definition of the interaction energy. Moreover, the very definition of
the interaction energy involves black holes, in contrast with previous
calculations where the black holes appear indirectly.

The interaction energy defined by Eq. (\ref{eq:1}) uses the fact that
we are choosing a particular topology for the initial data, but other
topologies are possible. Examples of different kind of topologies
where we can not use Eq. (\ref{eq:1}) are the Misner topology of two
isometric sheets \cite{Misner63},  the Misner wormhole
\cite{Misner60} or even initial data with trivial topologies which
contain an apparent horizon \cite{Beig91c}. If the initial data have
$k$ disconnected apparent horizons of area $A_k$, we can define the
individual masses as follows
\begin{equation}
  \label{eq:48}
 M_{H_k}=\sqrt{\frac{A_k}{16\pi}}. 
\end{equation}
Then, the interaction energy is given by the formula
\begin{equation}
  \label{eq:1b}
  E_{H}=M-\sum^N_{\substack{k'=0\\ k'\neq k}}M_{{H_k'}}.
\end{equation}
What is the relation between $E_k$ defined by Eq. (\ref{eq:1}) and
$E_{H}$ defined by (\ref{eq:1b})? Note that $E_{H}$ is much harder to
compute then $E_k$. I want to argue that $E_k$ will presumably give the
same result as $E_k$ for the leading orders terms in the expansion
given by theorem \ref{T1}. Assume that the data is such that when the
separation distance parameter is large, then the areas can by
approximated by the Schwarzschild formula $A_k\approx 16\pi M^2_k$ plus 
terms of order $J^2$. Then the
radius of the horizons will be $R_k\approx 2M_k$. The distance $l_{\tilde h}$
will differ from $l$ in terms order $O(M,J^2)$. Then we can replace
$l$ by $l_{\tilde h}$ in theorem $l_{\tilde h}$ and $M_{H_k}$ by $M_k$
in Eq. (\ref{eq:1}), up to this order.

It is interesting to note that the interaction energy $E_H$ has been
used in a different context, namely to determine the last stable circular
orbit in a black hole collision \cite{Cook94}.

\section*{Acknowledgments}
I would like to thank J. Isenberg and J. Valiente-Kroon for their
reading of the first draft of the paper and L. Szabados for discussions 
concerning quasilocal angular momentum. I would also like to thank the 
friendly hospitality of American Institute of Mathematics (AIM), where 
part of this work was done. 


\end{document}